\newif\ifpreprint

\documentclass[aps,prl,twocolumn,superscriptaddress]{revtex4}

\usepackage{amsmath,amsfonts}
\usepackage{epsfig}
\usepackage[breaklinks=true,bookmarks=true,hyperfigures=true]{hyperref}

\DeclareMathOperator{\Li}{Li}
\newcommand{\beq}{\begin{equation}}
\newcommand{\eeq}{\end{equation}}
\newcommand{\beqn}{\begin{eqnarray}}
\newcommand{\eeqn}{\end{eqnarray}}
\newcommand{\beqs}{\begin{eqnarray*}}
\newcommand{\eeqs}{\end{eqnarray*}}

\begin{document}

\preprint{DESY 10-197}

\title{ MHV amplitude for $3\to3$ gluon scattering in Regge limit}
\author{J.~Bartels}
\affiliation{II. Institute of  Theoretical Physics, Hamburg University, Germany}
\author{L.~N.~Lipatov}
\affiliation{II. Institute of  Theoretical Physics, Hamburg University, Germany}
\affiliation{St. Petersburg Nuclear Physics Institute, Russia}
\author{A.~Prygarin}
\affiliation{II. Institute of  Theoretical Physics, Hamburg University, Germany}

\begin{abstract}
We calculate  corrections to the BDS formula for the six-particle planar MHV amplitude for the gluon transition $3 \to 3$ in the multi-Regge kinematics for the physical region, in which the Regge pole ansatz is not valid.  The remainder function at two loops is obtained by an analytic continuation of the expression derived by Goncharov, Spradlin, Vergu and Volovich to the kinematic region described by the Mandelstam singularity exchange in the crossing channel. It contains both the imaginary and real contributions being in agreement with the BFKL predictions. The real part of the three loop expression is found from a dispersion-like all-loop formula for   the remainder function in the multi-Regge kinematics derived by one of the authors. We also make a prediction for the all-loop real part of the remainder function multiplied by the BDS phase, which can be accessible through  calculations in the regime of the strong coupling constant.
\end{abstract}

\maketitle

\section{Introduction}
In recent years we witnessed a significant progress in understanding  the structure of the scattering amplitudes in the supersymmetric theories. The pioneering paper of Parke and Taylor~\cite{Parke:1986gb} on the MHV amplitudes eventually led to a formulation of a simple all-loop expression for multi-leg amplitudes in $\mathcal{N}=4$ SYM by Anastasiou, Bern, Dixon and  Kosower~(ABDK)~\cite{Anastasiou:2003kj} and then by Bern, Dixon and Smirnov~(BDS)~\cite{Bern:2005iz}. However it was shown by two of the authors of this study in collaboration with Sabio Vera~\cite{BLS1} that the BDS ansatz is violated at two loops starting from six external gluons, confirming a conclusion derived by Alday and Maldacena~\cite{Alday:2007he} that the BDS formula is to be violated at large number of external gluons. It was argued by two of the authors~\cite{BLS1} that this violation is related to the fact that the BDS amplitude is not compatible with the Steinmann relations~\cite{Steinmann}, imposing the absence of simultaneous singularities in the overlapping channels.
Moreover, the BDS ansatz in some channels does not contain the contributions of the so-called Mandelstam cuts, which are the moving Regge singularities in the complex momenta plane. We call these channels the Mandelstam channels. The analytic properties of the BDS amplitude in the Regge kinematics were also investigated  by Brower, Nastase, Schnitzer and  C.-I Tan~\cite{Brower:2008nm,Brower:2008ia}.\newline
The BDS amplitude differs from the full MHV amplitude by a factor~\cite{Alday:2007hr} being a function  of the dual conformal invariants according to the analysis of Drummond, Henn, Korchemsky and Sokatchev~\cite{Drummond:2007au}. This function is commonly referred to as the remainder function $R_n^{(l)}$ for the $n$ external legs at $l$ loops.
The leading logarithmic term of $R_6^{(2)}$ for the Mandelstam channels of the $2\to 4$ and $3 \to 3$ scattering amplitudes in the Regge kinematics was explicitly calculated by the authors of ref.~\cite{BLS2} using a solution to the color octet Balitsky-Fadin-Kuraev-Lipatov~(BFKL)~(\cite{BFKL}) equation, which is a special case of the Schr\"odinger equation for the  open integrable Heisenberg spin chain~\cite{Lipatov:2009nt}. \newline
 It was suggested  that  in general kinematics $R_n^{(l)}$ can be obtained from the expectation value of the light-like polygonal Wilson loops~\cite{Alday:2007hr}. In particular, the remainder function for the six-gluon MHV amplitude at two loops was calculated by Drummond, Henn, Korchemsky and Sokatchev~\cite{Drummond:2007bm} and then it was expressed in terms of the generalized polylogarithms by Del Duca, Duhr and Smirnov~\cite{DelDuca:2009au,DelDuca:2010zg}. Their lengthy expression for $R^{(2)}_{6}$ was greatly simplified by Goncharov, Spradlin, Vergu and Volovich~(GSVV)~\cite{Goncharov:2010jf} and written in terms of only  classical polylogarithms. The GSVV expression was analytically continued by two of us~\cite{LP1}  to the Mandelstam channel of the $2 \to 4$ scattering amplitude considered in ref.~\cite{BLS2} and showed a full agreement within  leading logarithmic accuracy. The leading logarithmic term  and the real part of the next-to-leading term of the remainder function at three loops were found in ref.~\cite{LP2}. The analytic continuation at the strong coupling was performed by one of the authors with collaborators~\cite{Bartels:2010ej}.     \newline
In the present paper we consider $3 \to 3$ gluon scattering amplitude in the Mandelstam  channels. This amplitude  is generally not related to the $2\to4 $ amplitude considered in the previous studies~\cite{LP1,LP2} and brings new information about the analytic structure of the six-gluon MHV amplitude.  We perform the analytic continuation  of the GSVV expression for the two-loop remainder function to the Mandelstam  channel of the $3\to 3$ amplitude and simplify it in the Regge limit. The result is similar to that of the $2 \to 4$ case and  differs by the overall sign and the presence of the real contribution. The obtained real contribution confirms general all-loop dispersion relations for the real and imaginary parts of the remainder function derived by one of the authors~\cite{Lipatov:2010qf}. These dispersion relations are used to calculate also the  leading logarithmic terms and the real part of the next-to-leading contribution at three loops. Another important result of the present study is the prediction, in the region under consideration, of the real constant part of the remainder function multiplied by the phase present in the BDS amplitude. This prediction is valid for an arbitrary value of the coupling constant and can be accessible through the strong coupling calculations.

\section{\label{sec:remainder-function}Analytic continuation}

\begin{figure}[htbp]
	\begin{center}
		\epsfig{figure=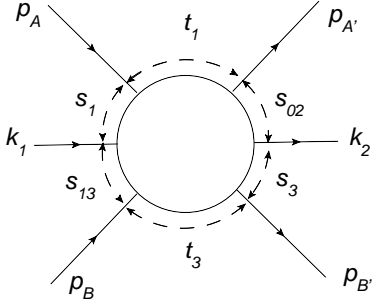,width=50mm}
	\end{center}
	\caption{ The $3 \to 3$ gluon scattering amplitude. }
	\label{fig:3to3direct}
\end{figure}

\begin{figure}[htbp]
	\begin{center}
		\epsfig{figure=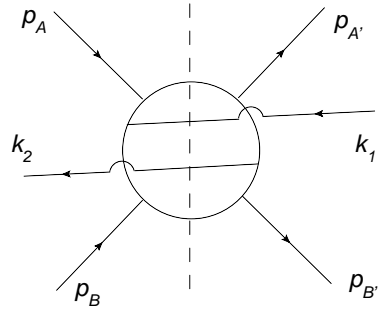,width=50mm}
	\end{center}
	\caption{ The $3 \to 3$ gluon scattering amplitude in the Mandelstam  channel given by $s_1, s_3, s_{13}, s_{02} <0\;\;\; \text{and} \;\;\; s, t^{'}_2>0$.  }
	\label{fig:3to3NOTdirect}
\end{figure}

We consider a special case of the six-gluon planar MHV scattering amplitude for three gluon scattering~($3 \to 3$ amplitude) illustrated in Fig.~\ref{fig:3to3direct}.
The energy invariants  are defined by  $s_{13}=(p_B+k_1)^2,  s_{02}=(p_{A'}+k_2)^2,
s=(p_B+k_1+p_A)^2, t^{'}_{2}=(p_A-p_{A'}-k_2)^2, s_1=(k_1+p_A)^2,  s_3=(p_{B'}+k_2)^2, t_2=(p_{A}-p_{A'}+k_1)^2, t_1=(p_A-p_{A'})^2$ and $t_3=(p_{B}-p_{B'})^2$. The dual conformal cross ratios are expressed in terms of the energy invariants as follows
\beqn\label{crossinv}
u_1=\frac{s_{13}s_{02}}{s\; t^{'}_2},\; u_2=\frac{s_{1}s_{3}}{s\; t_2}, \; u_3=\frac{t_{1}t_{3}}{t_2  t^{'}_2}.
\eeqn
In the multi-Regge  kinematics for  the direct channel, where all invariants are negative
\beqn
-s \gg -s_1, -s_3, -t^{'}_2 \gg -t_1, -t_2, -t_3 >0,
\eeqn
 the remainder function $R_6^{(l)}$ is zero, while in the physical region of the Mandelstam  channel depicted in Fig.~\ref{fig:3to3NOTdirect}, where
\beqn\label{mandchannel}
s_1, s_3, s_{13}, s_{02} <0\;\;\; \text{and} \;\;\; s, t^{'}_2>0,
\eeqn
it contains a non-vanishing contribution.

This situation was thoroughly  discussed in refs.~\cite{BLS1,BLS2} as well as in ref.~\cite{Lipatov:2010qf}. The physical reason for the violation of the BDS ansatz in this region is the fact that the BDS formula does not have correct analytic properties, in particular, it does not account properly for the Mandelstam~(Regge) cuts.
In the Mandelstam  channel~(\ref{mandchannel}) in the multi-Regge kinematics the dual conformal cross ratios (\ref{crossinv}) possess a non-zero phase
\beqn\label{3to3cont}
u_1 \to |u_1| e^{i 2\pi}, \; u_2 \to |u_2| e^{i \pi},\; u_3 \to |u_3| e^{i \pi}.
\eeqn
Using this phase structure  one can perform an analytic continuation of the remainder function to our kinematic region.
The two-loop remainder function for the six-gluon MHV amplitude in terms of the classical polylogarithms was calculated by Goncharov, Spradlin, Vergu and Volovich~\cite{Goncharov:2010jf}. They found that in the variables
\begin{equation}
x^\pm_i = u_i x^\pm, \qquad
x^\pm = \frac{u_1+u_2+u_3-1 \pm \sqrt{\Delta}}{2 u_1 u_2 u_3},
\end{equation}
where $\Delta = (u_1+u_2+u_3-1)^2 - 4 u_1u_2u_3$,
the remainder function $R_6^{(2)}$ can be written in a rather compact way
\begin{multline}
\label{eq:mainresult}
R^{(2)}_6(u_1,u_2,u_3) = \sum_{i=1}^3 \left( L_4(x^+_i, x^-_i) -
\frac{1}{2} \Li_4(1 - 1/u_i)\right) \cr
- \frac{1}{8} \left( \sum_{i=1}^3 \Li_2(1 - 1/u_i) \right)^2
+ \frac{1}{24}J^4 +  \frac{\pi^2}{12} J^2 + \frac{\pi^4}{72}.
\end{multline}
The functions $L_4(x^+, x^-)$  and $J $ are defined by
\begin{multline}
\label{eq:bwrz}
L_4(x^+, x^-) =
\frac{1}{8!!} \log(x^+ x^-)^4
\cr
+
\sum_{m=0}^3
\frac{(-1)^m}{(2m)!!} \log(x^+ x^-)^m
(\ell_{4-m}(x^+) + \ell_{4-m}(x^-))
\end{multline}
and
\beqn
&& \ell_n(x) = \frac{1}{2} \left( \Li_n(x) - (-1)^n \Li_n(1/x) \right),
\\ &&
J = \sum_{i=1}^3 (\ell_1(x^+_i) - \ell_1(x^-_i))\nonumber.
\eeqn

\begin{figure}[htbp]
	\begin{center}
		\epsfig{figure=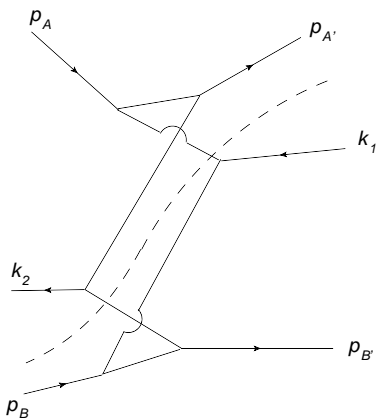,width=50mm}
	\end{center}
	\caption{ The $3 \to 3$ gluon scattering amplitude with discontinuity in $t^{'}_2$ channel.}
	\label{fig:2to4rot}
\end{figure}

In this paper we perform the analytic continuation of (\ref{eq:mainresult}) in phases of  $u_i$ given by  (\ref{3to3cont}) to the Mandelstam  channel of the $3 \to 3$ scattering amplitude in the multi-Regge limit.
We find that it is not so much different from the remainder function in the Mandelstam  channel of the $2\to 4$ gluon scattering amplitude calculated by the authors in ref.~\cite{LP1,LP2}. This fact can be explained by the Regge factorization as discussed below. The main difference between the two is that in the $3 \to 3$ case we get a non-vanishing real contribution in the next-to-leading logarithmic approximation~(NLLA) as predicted by one of the authors~\cite{Lipatov:2010qf}. At two loops we keep only the leading order~(LLA) and constant terms~(NLLA) in the logarithm of the energy $\ln(u_1-1)$ \footnote{Note that in the $3\to3$ case $u_1>1$, in contrast to the $2 \to 4$ case, where $u_1<1$. For more details the reader is referred to refs.~\cite{BLS1,BLS2}.}, where
\beqn
u_1 -1 \simeq \frac{(\mathbf{q}_1+\mathbf{q}_3-\mathbf{q}_2)^2}{|t^{'}_{2}|}.
\eeqn
The logarithmic contributions in $\ln(u_1-1)$ originate from the discontinuities in $s$- and $t^{'}_2$-channels illustrated in Fig.~\ref{fig:3to3NOTdirect} and Fig.~\ref{fig:2to4rot} respectively. Each order of the perturbation theory brings a power of $\ln(u_1-1)$ in the multi-Regge kinematics, so that at two loops one expects at most the first power of $\ln(u_1-1)$ (we start with an imaginary constant at one loop) and at three loops there appears a term proportional to $\ln^2(u_1-1)$.

The remainder function after the analytic continuation (\ref{3to3cont}) to the Mandelstam  channel of $3 \to 3 $ amplitude in the multi-Regge is given by
\beqn\label{R62cont3to3}
&& R_{6}^{(2)}\left(|u_1|e^{i2\pi}, \frac{e^{i\pi}|1-u_1|}{|1+w|^2},  \frac{e^{i\pi}|1-u_1||w|^2}{|1+w|^2}\right)
\\
&& \simeq -\frac{i\pi}{2}\ln(u_1-1)\ln|1+w|^2 \ln \left|1+\frac{1}{w}\right|^2 \nonumber
\\
&&   +\frac{\pi^2}{2}\ln|1+w|^2 \ln \left|1+\frac{1}{w}\right|^2   -\frac{i\pi}{2}\ln|w|^2 \ln^2|1+w|^2
\nonumber\\
&& +\frac{i\pi}{3}\ln^3|1+w|^2 -i\pi \ln |w|^2 \left(\text{Li}_2(-w)+\text{Li}_2(-w^*)\right) \nonumber\\
&&
+i2\pi \left(\text{Li}_3(-w)+\text{Li}_3(-w^*)\right). \nonumber
\eeqn
The complex variable $w$ is expressed in terms of the reduced cross ratios
\beqn
\tilde{u}_2=\frac{|u_2|}{|1-u_1|},\; \tilde{u}_3=\frac{|u_3|}{|1-u_1|}
\eeqn
through
\beqn
w=\frac{B^{+}}{\tilde{u}_2}, \; w^*=\frac{B^{-}}{\tilde{u}_2}
\eeqn
for $B^{\pm}$ defined by~\cite{LP2}
\beqn
B^{\pm}=\frac{1-\tilde{u}_2-\tilde{u}_3 \pm \sqrt{(1-\tilde{u}_2-\tilde{u}_3)^2-4 \tilde{u}_2 \tilde{u}_3}}{2}.
\eeqn
 In the course of the analytic continuation we obtain  terms of the order  $\ln^2(u_1-1)$ and $\ln^3(u_1-1)$. These higher order terms in the logarithm of the energy  all cancel in the final result. The remainder function $R_{6}^{(2)}$ in (\ref{R62cont3to3}) is symmetric under substitution  $w \to 1/w$ and vanishes in the limit $|w| \to 0$ or $|w| \to \infty$.

The expression in (\ref{R62cont3to3}) was obtained by the analytic continuation for the $3\to 3 $ amplitude (see (\ref{3to3cont})) of the remainder function of Goncharov et al. in  (\ref{eq:mainresult}) and then simplified in the Regge limit
\beqn\label{3to3regge}
|u_1| \to 1^{+}, \; |u_2| \to 0^+,\; |u_3| \to 0^+, \; \tilde{u}_{2,3} \sim \mathcal{O}(1).
\eeqn
However it is also possible to get the same expression from the remainder function for the $2 \to 4$ amplitude with the use of its cyclic symmetry~\cite{Lipatov:2010qf}.
The remainder function for the $2 \to 4$ case was found by two of the authors~\cite{LP1,LP2} analytically continuing (\ref{eq:mainresult}) in
\beqn\label{2to4cont}
u_1 \to |u_1| e^{-i 2\pi}, \; u_2 \to u_2 ,\; u_3 \to u_3
\eeqn
and then simplifying in the Regge limit
\beqn\label{2to4regge}
|u_1| \to 1^{-}, \; |u_2| \to 0^+,\; |u_3| \to 0^+, \; \tilde{u}_{2,3} \sim \mathcal{O}(1).
\eeqn

\begin{figure}[htbp]
	\begin{center}
		\epsfig{figure=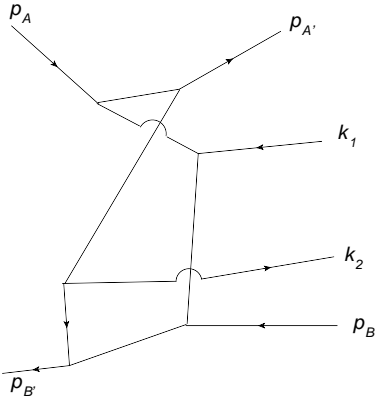,width=50mm}
	\end{center}
	\caption{ The $3 \to 3$ gluon scattering amplitude for the Mandelstam channel, where $t_2', s, s_1, s_3 < 0$ and $s_{13}, s_{02} >0$. }
	\label{fig:2to4rotNew}
\end{figure}

Another interesting physical region  of the Mandelstam channel is depicted in Fig.~\ref{fig:2to4rotNew}. It can be obtained from  Fig.~\ref{fig:2to4rot} by twisting the lower part of figure.  In this region $t_2', s, s_1, s_3 < 0$ and $s_{13}, s_{02} >0$, so that the
corresponding analytic continuation is given by
\beqn\label{2to4contNew}
u_1 \to |u_1| e^{-i 2\pi}, \; u_2 \to u_2 ,\; u_3 \to u_3.
\eeqn
Thus this case is trivially related to the $2 \to 4$ amplitude considered in refs.~~\cite{BLS1,BLS2,LP1,LP2}, in particular, here we also have $|u_1|<1$. The similarity between the two cases is expected from the Regge factorization of the scattering amplitudes.  In the rest of the present paper we focus  on the non-trivial region of the $3 \to 3$ amplitude shown in Fig.~\ref{fig:2to4rot}.

The remainder function for the Mandelstam  channel of the $3\to 3$ case in (\ref{R62cont3to3}) differs from the remainder function for the corresponding  channel of the $2 \to 4$ case (see (22) of ref.~\cite{LP2}) only by the overall sign and the presence of the real term subleading in the logarithm of the energy $\ln(u_1-1)$. The origin of this difference is best understood from a general expression for remainder function both for $3\to3$ and $2 \to 4$ cases derived by one of the authors~\cite{Lipatov:2010qf}.
For the $2 \to 4$ amplitude in region defined by (\ref{2to4cont}) and (\ref{2to4regge}) it is given by
\beqn\label{Reqn2to4}
R_6 e^{i\pi \delta} =\cos \pi \omega_{ab} +i  \int_{-i\infty}^{i\infty} \frac{d \omega }{2\pi i}f(\omega) e^{-i\pi \omega}|1-u_1|^{-\omega}, \;\;\;\;
\eeqn
and for the $3 \to 3$ amplitude in region defined by (\ref{3to3cont}) and (\ref{3to3regge})  it reads
\beqn\label{Reqn3to3}
R_6 e^{-i\pi \delta} =\cos \pi \omega_{ab} -i  \int_{-i\infty}^{i\infty} \frac{d \omega }{2\pi i}f(\omega) |1-u_1|^{-\omega}, \;\;\;\;
\eeqn
where $R_{6}$ are the remainder functions for  corresponding process.
The phases $\delta$ and $\omega_{ab}$ are defined by~\cite{Lipatov:2010qf}
\beqn\label{deltaomega}
&& \delta= \frac{\gamma_K}{8}\ln \tilde{u}_2 \tilde{u}_3=\frac{\gamma_K}{8} \ln \frac{|w|^2}{|1+w|^4}
,\\
&& \; \omega_{ab}=\frac{\gamma_K}{8}\ln \frac{\tilde{u}_3}{\tilde{u}_2}=\frac{\gamma_K}{8} \ln |w|^2, \nonumber
\eeqn
where $\gamma_K$ is the cusp anomalous dimension $\gamma_K \simeq 4 a$ for $a=g^2 N_c/8/\pi^2$ and  $\omega $ is related to the angular momentum in $t_2$-channel~\footnote{More details on the equations (\ref{Reqn2to4}) and (\ref{Reqn3to3}), including the rigorous definitions of $\omega$ and $f(\omega)$ are presented in ref.~\cite{Lipatov:2010qf,LP2}}.

The all-loop expressions in (\ref{Reqn2to4}) and (\ref{Reqn3to3}) have a meaning of dispersion relations, which establish a connection between real and imaginary parts of the scattering amplitude.
 It is worth emphasizing that the integral term in (\ref{Reqn2to4}) and (\ref{Reqn3to3}) is formally divergent at one loop and should be understood in the sense of the principal value prescription~(cf.~\cite{BLS2}). It cancels the one-loop contribution from the BDS phase $\delta$, so that $R_6^{(1)}$ is zero as expected.

 A few words to be said about the structure of (\ref{Reqn2to4}) and (\ref{Reqn3to3}). As it was discussed in refs.~\cite{BLS1,Lipatov:2010qf} assuming the Regge pole factorization the six-gluon amplitude can be written as five
 contributions compatible with the Steinmann relations. Values of four out of five relative coefficients are fixed by
  the BDS amplitude in the four physical regions. Using the Weis factorization property~\cite{Weis:1972ir}
  one can fix the whole Regge pole structure of the six-gluon amplitude, however the resulting expression has unpleasant analytic properties. Namely, it includes  some singularities incompatible with the perturbation theory.  It was argued by one of the authors~\cite{Lipatov:2010qf} that these dangerous terms can be absorbed in the Mandelstam cut contribution because they have the same phase structure. The resulting expressions have correct analytic properties and can be written in the form of (\ref{Reqn2to4}) and (\ref{Reqn3to3}). The contributions of  the Regge pole ($\cos \pi \omega_{ab}$) and the Mandelstam cut~(the integral over $\omega$) are functions of only dual conformal cross ratios. The factors  $e^{\pm i\pi \delta}$ for the corresponding physical regions accounts for a phase already present in the BDS amplitude. They are extracted from the BDS amplitude to make it self-consistent. For more details the reader  is referred to section 2 of ref.~\cite{Lipatov:2010qf}. The "dispersion" relations in (\ref{Reqn2to4}) and (\ref{Reqn3to3}) are correct in the Regge kinematics for any number of loops and thus allow us to make predictions also for a strong coupling regime as discussed below.

Substituting the expansion of  the remainder function
 \beqn
 R_6 \simeq 1+a^2R^{(2)}_6
 \eeqn
and the leading logarithmic~(LLA) approximation of $R^{(2)}_6$ for $2 \to 4$ amplitude calculated in ref.~\cite{BLS2}
we find that the subleading in $\ln(1-u_1)$ real term in $R^{(2)}_6$ fully cancels with the contributions from $\delta$ and $\omega_{ab}$~\cite{Lipatov:2010qf}. The physical meaning of this is that contributions from Regge poles (the cosine term depending on $ \omega_{ab}$ )  and Mandelstam cuts with the factor $e^{-i\pi \omega}$ in the integral of (\ref{Reqn2to4}) are related to each other due to the analyticity of the amplitude.

This cancellation does not happen for the  $3 \to 3$ amplitude, where the contribution from the Mandelstam cut is pure imaginary (no  factor $e^{-i\pi \omega}$ in the integrand). The contribution from Regge poles $\cos \pi \omega_{ab}$ is pure real, but the phase of BDS amplitude $\delta$ mixes between real and imaginary parts of the remainder function. Thus the $3 \to 3$  remainder function does have a real part at two loops in region (\ref{3to3cont}). Indeed, expanding (\ref{Reqn3to3}) to the second order in $a$ we obtain
\beqn\label{3to3twoloops}
&& a^2 R^{(2)}_6 -\frac{\pi^2 \delta^2}{2} =-\frac{\pi^2 \omega^2_{ab}}{2}
\\
&&-i\frac{a^2}{2}\frac{\partial^2}{\partial a^2}\left( \int_{-i\infty}^{i\infty} \frac{d \omega }{2\pi i}f(\omega) (u_1-1)^{-\omega}\right)|_{a=0}. \;\;\;\; \nonumber
\eeqn
The integral term in (\ref{3to3twoloops}) gives a pure imaginary contribution and thus the real part of the $3 \to 3$ remainder function reads
\beqn
&&\Re \left(R^{(2)}_6\right)= \frac{\pi^2 \delta^2}{2 a^2} -\frac{\pi^2 \omega^2_{ab}}{2 a^2}=\frac{\pi^2}{2}\ln \tilde{u}_3 \ln \tilde{u}_2 \\
&&=\frac{\pi^2}{2}\ln |1+w|^2 \ln \left|1+\frac{1}{w}\right|^2, \nonumber
\eeqn
in full agreement with (\ref{R62cont3to3}).

From (\ref{Reqn2to4}) and (\ref{Reqn3to3}) we deduce that the remainder function of the $3 \to 3$ amplitude can be obtained from that of the $2 \to 4 $ amplitude by a simple transformation
\beqn\label{trans}
\ln(1-u_1) \to \ln(u_1-1)-i\pi
\eeqn
together with the subsequent complex conjugation.
This is related to the fact, that the  Mandelstam cut contribution is  constructed from impact factors and the corresponding Green functions, which are the same for $2 \to 4$ and $3 \to 3$ amplitudes.
It is easy to see that the transformation (\ref{trans})  is true for (22) of ref.~\cite{LP2} and (\ref{R62cont3to3}).

\section{ $3 \to 3$ remainder function at three loops}
In this section we find the remainder function for the $3 \to 3$ scattering amplitude at three loops with logarithmic accuracy  $R_6^{(3)\; LLA}$ as well as  the real part of the  next-to-leading logarithmic term $\Re \left(R_6^{(3)\; NLLA}\right)$.
In accordance to the general analytic properties of the scattering amplitudes the leading term in the leading logarithmic approximation~(LLA) is pure imaginary. The integral all-order representation for the LLA part of the remainder function was found in ref.~\cite{BLS2} using the solution to the octet BFKL equation. It was shown that the LLA term of the remainder function at arbitrary numbers of loops in the Mandelstam  channel of the $3 \to 3$ amplitude differs from that of the $2 \to 4$ amplitude only by the overall sign. The explicit expression for $R_6^{(3)\; LLA}$ in the Mandelstam  channel of the $2\to 4$ amplitude was found in ref.~\cite{LP2}. Using this result we readily calculate the leading logarithmic contribution to the remainder function of the Mandelstam  channel of the $3 \to 3$ amplitude at three loops
\beqn\label{R63LLA3to3}
&& \hspace{-0.5cm}R_6^{(3)\; LLA}=-\frac{i\pi}{4}\ln(u_1-1)^2\left(
\ln|w|^2 \ln^2|1+w|^2
\right. \hspace{0.5cm}\\
&& \left.
-\frac{2}{3}\ln^3|1+w|^2
+\frac{1}{2} \ln|w|^2 \left(\text{Li}_2 (-w)+\text{Li}_2 (-w^*)\right)
\right.\nonumber\\
&& \left.
-\frac{1}{4}\ln^2|w|^2 \ln|1+w|^2-\text{Li}_3 (-w)-\text{Li}_3 (-w^*)
\right).\nonumber
\eeqn
The real part of the next-to-leading contribution is found  expanding (\ref{Reqn3to3}) to the third order in  $a$ and extracting the real terms
\beqn\label{real2l}
\Re \left(R_6^{(3)\; NLLA}\right) = \frac{i\pi \delta }{a} R_6^{(2)\; LLA}.
\eeqn
For an arbitrary number of loops $\ell \geq 3$ this reads
\beqn
  \Re \left(R_6^{(\ell)\; NLLA}\right) =\frac{i\pi \delta}{a} R_6^{(\ell-1)\; LLA},
\eeqn
where
\beqn
R_6^{(\ell)\; LLA} =\frac{-i}{\ell!} \frac{\partial^{\ell} }{\partial a^{\ell}}  \left (\int_{-i\infty}^{i\infty} \frac{d \omega }{2\pi i}f(\omega) (u_1-1)^{-\omega}\right)|_{a=0} \;\;
\eeqn
as follows from (\ref{Reqn3to3}).

Using $R_6^{(2)\; LLA}$ given by  the first term on RHS of (\ref{R62cont3to3}) we get
\beqn\label{R63reNLLA3to3}
&& \hspace{-0.5cm}\Re \left(R_6^{(3)\; NLLA} \right)=\frac{\pi^2}{4}\ln(u_1-1)\ln \left(\tilde{u}_2 \tilde{u}_3\right)\ln \tilde{u}_2\ln \tilde{u}_3 \nonumber\\
&&=-\frac{\pi^2}{4}\ln(u_1-1)\left( \ln^2 |1+w|^2 \ln \left|1+\frac{1}{w} \right|^2\right.\;\;\;\nonumber \hspace{-1cm}\\
&&
\left.
 +\ln |1+w|^2 \ln^2 \left|1+\frac{1}{w} \right|^2 \right).
\eeqn
The full remainder function at three loops in the Regge kinematics has also NLLA imaginary and next-to-next-to-leading logarithmic~(NNLLA) contributions. To find them one needs to know the higher order corrections to the function $f(\omega)$ in (\ref{Reqn3to3}), which are not available at the moment apart from the next-to-leading contribution to the impact factor calculated in ref.~\cite{LP2}.

Both $R_6^{(3)\; LLA}$ and $R_6^{(3)\; NLLA}$ are symmetric under inversion $w \to 1/w$ and vanishing for $|w|\to 0$ or $|w|\to \infty$.
The $w \to 1/w$ symmetry implies a target-projectile symmetry, where the amplitude is symmetric with respect to the transformation $p_{A},p_{A'},p_{B},p_{B'},k_1,k_2 \to p_{B},p_{B'},p_{A},p_{A'},k_2,k_1$.

As it was already mentioned the BDS phase $\delta$ mixes between real and imaginary parts of the remainder function. This means that by virtue of (\ref{Reqn3to3}) the real part of the remainder function at an arbitrary number of loops $R_6^{(l)}$ is expressed through the BDS phase $\delta$ and the  remainder function with  a lower number of loops $R_6^{(l-1)}$, $R_6^{(l-2)}$ etc. Despite this fact we can make an all-loop prediction for a value of
\beqn\label{allrealsub}
\Re\left(R_{6}e^{-i\pi \delta}\right)= \cos \pi \omega_{ab},
\eeqn
where $\delta$ and $\omega_{ab}$ are given by (\ref{deltaomega}). In the Regge limit $\Re\left(R_{6}e^{-i\pi \delta}\right)$ gives the constant term, which is not accompanied by any logarithm of the energy $\ln(u_1-1)$.
This all-loop result can be accessible through  calculations in the strong coupling regime. However there is some difficulty in understanding (\ref{allrealsub}) in this regime related to the fact that at large coupling constants the functions $\delta$ and $\omega_{ab}$  grow, and therefore the expression in (\ref{allrealsub}) rapidly oscillates and does not have a definite limit.  Note, that the expression in (\ref{allrealsub}) is valid only for the Mandelstam  channel of the  $3 \to 3$ scattering amplitude in the multi-Regge kinematics. In  the $2 \to 4 $ case this simple structure is spoiled by the presence of the $e^{-i\pi \omega}$ factor in the integral in (\ref{Reqn2to4}), although this factor disappears after the analytic continuation to the non-physical region $u_1>1$.

\section{\label{sec:discussion}Discussion}
In the present study we consider $3\to 3$ planar gluon MHV amplitude in the multi-Regge kinematics. We perform the analytic continuation of the six-gluon remainder function at two loops found by Goncharov, Spradlin, Vergu and Volovich to the Mandelstam  channel illustrated in Fig.~\ref{fig:3to3NOTdirect} and then extract the logarithmic and constant terms in the Regge limit. We find that despite the fact that $2 \to 4$
and $3 \to 3$ amplitudes have a rather different structure, the corresponding remainder functions have a similar form as expected from the Regge factorization of  scattering amplitudes. The only difference between them at two loops is the overall sign and the presence of the real term for the $3 \to 3$ remainder function. This result is in full agreement with a general all-loop dispersion relations (\ref{Reqn2to4}) and (\ref{Reqn3to3}) derived by one of the authors~\cite{Lipatov:2010qf}. Using these dispersion relations  we predict the leading term~(\ref{R63LLA3to3}) and the subleading real term~(\ref{R63reNLLA3to3}) of the $3 \to 3$ remainder function at three loops. We also make a prediction for all-loop expression of the real part of the $3 \to 3$ remainder function multiplied by the BDS phase~(\ref{allrealsub}). This relation can be accessible through  calculations in the regime of the large coupling constant.

\section*{Acknowledgments}

We are deeply indebted to I.~Balitsky, J.~Kotanski, A.~Mueller, A.~Sabio Vera, V.~Schomerus,  A.~Sever, M.~Spradlin,
C.~Vergu,  P.~Vieira and A.~Volovich    for fruitful discussions and suggestive comments.

\end{document}